\pdfoutput=1

\documentclass[aps,prb,twocolumn,superscriptaddress]{revtex4}
\usepackage{graphicx} 
\usepackage{dcolumn}
\usepackage{bm}

\begin{document}

\title{High-Ampacity Power Cables of Tightly-Packed and Aligned Carbon Nanotubes}

\author{X.~Wang}
\affiliation{Department of Electrical and Computer Engineering, Rice University, Houston, TX 77005, USA}
\affiliation{The Richard E. Smalley Institute for Nanoscale Science and Technology, Rice University, Houston, TX 77005, USA}
\author{N.~Behabtu}
\author{Colin C. Young}
\author{Dmitri E. Tsentalovich}
\affiliation{The Richard E. Smalley Institute for Nanoscale Science and Technology, Rice University, Houston, TX 77005, USA}
\affiliation{Department of Chemical and Biomolecular Engineering, Rice University, Houston, TX 77005, USA}
\author{M.~Pasquali}
\affiliation{The Richard E. Smalley Institute for Nanoscale Science and Technology, Rice University, Houston, TX 77005, USA}
\affiliation{Department of Chemical and Biomolecular Engineering, Rice University, Houston, TX 77005, USA}
\affiliation{Department of Chemistry, Rice University, Houston, TX 77005, USA}
\affiliation{Department of Materials Science and NanoEngineering, Rice University, Houston, TX 77005, USA}
\author{J.~Kono}
\email[]{kono@rice.edu}
\thanks{corresponding author.}
\affiliation{Department of Electrical and Computer Engineering, Rice University, Houston, TX 77005, USA}
\affiliation{The Richard E. Smalley Institute for Nanoscale Science and Technology, Rice University, Houston, TX 77005, USA}
\affiliation{Department of Materials Science and NanoEngineering, Rice University, Houston, TX 77005, USA}
\affiliation{Department of Physics and Astronomy, Rice University, Houston, TX 77005, USA}

\date{\today}

\begin{abstract}
We characterize the current-carrying capacity (CCC), or ampacity, of highly-conductive, light, and strong carbon nanotube (CNT) fibers by measuring their failure current density (FCD) and continuous current rating (CCR) values. 
We show, both experimentally and theoretically, that the CCC of these fibers is determined by the balance between current-induced Joule heating and heat exchange with the surroundings.  The measured FCD values of the fibers range from 10$^7$ to 10$^9$~A/m$^2$ and are generally higher than the previously reported values for aligned buckypapers, carbon fibers, and CNT fibers.  To our knowledge, this is the first time the CCR for a CNT fiber has been reported.   We demonstrate that the specific CCC (i.e., normalized by the linear mass density) of our CNT fibers are higher than those of copper.
\end{abstract}


\maketitle


\section{Introduction}

Advanced materials utilizing carbon nanotubes (CNTs) are emerging.  Recently, we reported a macroscopic fiber composed of tightly packed and well-aligned CNTs, which combines specific strength, stiffness, and thermal conductivity of carbon fibers with the specific electrical conductivity of metals (``specific'': normalized by the linear mass density).\cite{BehabtuetAl13Science}  These macroscopic CNT fibers hold the promise to replace traditional metals for many applications including making stronger and lighter power transmission cables or electronic interconnections,\cite{JaroszetAl11Nanoscale} as well as durable field emission or thermionic emission sources.\cite{MurrayetAl13APL,WeietAl07NL}  These applications require the fiber to operate under high current, which leads to natural questions about the fiber's ability to carry such a current without being damaged.  Traditionally, current carrying capacity (CCC), or often called ampacity, is used to quantify this ability.  CCC is defined as the maximum amount of current a cable (including any insulating layer) can carry before sustaining immediate or progressive damages; sometimes, it is more convenient to use the current density, especially when making comparisons among different types of cables.  Also, for weight-critical applications, for instance, in the aerospace industry, specific CCC (CCC normalized by the linear mass density) is usually considered.

Owing to the strong C-C bond, the CCC of individual CNTs can exceed 10$^{13}$~A/m$^2$ without damage by electromigration,\cite{Yaoetal00PRL,WeietAl01APL,CollinsetAl01PRL} which is 2 to 3 orders of magnitude greater than the electromigration limit of copper.\cite{HauderetAl01APL}  However, such superb CCC (limited by intrinsic optical phonon emission) becomes unapproachable when many CNTs are packed together to form a macroscopic CNT fiber or bundle.  The unavoidable inter-tube transport significantly increases the resistivity, and the resultant Joule heating at high current densities raises the temperature, inducing damages and ultimately breaking the fiber.  Thus, the competition between current-induced Joule heating and  cooling by thermal environments becomes the determinant of the CCC, as in metal cables; this competition scales with the volume-to-surface ratio, which increases with increasing cable diameter, making Joule heating progressively more problematic for larger diameter cables.

So far, the most widely studied case for CNT networks is their immediate breakdown (usually in seconds or less) when carrying high current.  The damage usually initiates around the hottest spot, particularly if associated with defects, kinks, or impurities.\cite{KurodaetAl05PRL,HuangetAl05PRL,MolhaveetAl06NL,WeietAl08APL}  The corresponding current limit can be defined as the failure current density (FCD), similar to the fuse current limit for metal cables.  On the other hand, to be used as a power cable, CNT wires must operate below a regulations-specified temperature called the ``operating temperature'' ($T_{\rm o}$) to avoid damaging its own insulation layer or other nearby accessories.  The corresponding current limit is defined as the continuous current rating (CCR).\cite{FinkBeaty06Book}  Since $T_{\rm o}$ cannot be high enough to cause any damages, CCR is always much lower than the failure current.  In contrast to metal power cables, whose CCR is well studied\cite{Anders97Book} and regulated,\cite{FinkBeaty06Book} so far no systematic study of these quantities for CNT wires is available.

Here, we determined both FCD and CCR for CNT fibers under various test conditions.  We first measured the FCDs of those fibers.  We monitored how the resistivity of the fiber under test evolved as a function of current density and found four distinct regimes.  The measured FCDs varied from 10$^7$ to 10$^9$~A/m$^2$, depending on the dimensions of those fibers and test conditions. In particular, the measured FCD in vacuum was much lower than in gases due to a lack of heat exchange by gases, while the measured FCD in air was smaller than in the other tested gases because of oxidation.  We then analyzed the heat exchange between CNT fibers and each type of gas and extracted the thermal conductance ($g$) between them. We proved that the heat exchange is governed by natural convection.  In addition, we showed that due to tight packing and good alignment, gas molecules do not penetrate the body of the CNT fibers.  When $g$ is known, in principle, any thermally determined CCC can be deduced if the corresponding temperature limit is given, and vice versa.  As an example, we determined CCR for infinitely long CNT fibers with an operating temperature of 363~K.

Based on these measurements and heat-exchange analysis, we were able to make a comprehensive comparison of CCC with other cables.  We showed that the FCD of our fibers is higher than previously reported carbon fibers and CNT fibers.  
We then compared these two parameters with a pure copper wire.  Both the FCD  and CCR of copper were still higher than CNT fibers mainly due to copper's lower resistivity.  However, when normalized by the mass density, both specific FC (SFC) and CCR of copper wire were lower than those of these lightweight CNT fibers.  Considering the fact that commercial transmission cables usually require extra reinforcement by a steel core because of copper's heavier weight and lower tensile strength, the combination of higher specific CCC and stronger mechanical strength of CNT fibers makes them promising candidates for transmission cables.

\section{METHODS}
\label{methods}

\subsection{Sample preparation}

CNT fibers were produced by wet spinning.\cite{BehabtuetAl13Science}  Purified CNTs were dissolved in chlorosulfonic acid at a concentration of 3~wt$\%$ and filtered to form a spinnable liquid crystal dope.\cite{DavisetAl09NN}  The dope was then extruded through a spinneret (65--130~$\mu$m in diameter for different diameter fibers) into a coagulant (acetone or water) to remove the acid. The forming filament was collected onto a winding drum with a linear velocity higher than the extrusion velocity to enhance the alignment. The produced fibers were further washed in water and dried in an oven at 115$^{\circ}$C.  Such fiber is called an acid-doped fiber. TGA shows that there still remains about 7 wt$ \% $ of acid residuals in the acid-doped fiber.\cite{BehabtuetAl13Science} On the other hand, if the produced fibers were first dried in an oven at 115$^{\circ}$C and then washed in water, there would be even more acid residuals in the fiber. Such fiber is called a heavily acid-doped fiber.
The 99.99\% pure copper wire with 0.001 inch diameter was purchased from ESPICorp, Inc.

\subsection{Experimental setup}

\begin{figure}
\begin{center}
\includegraphics[scale=0.55]{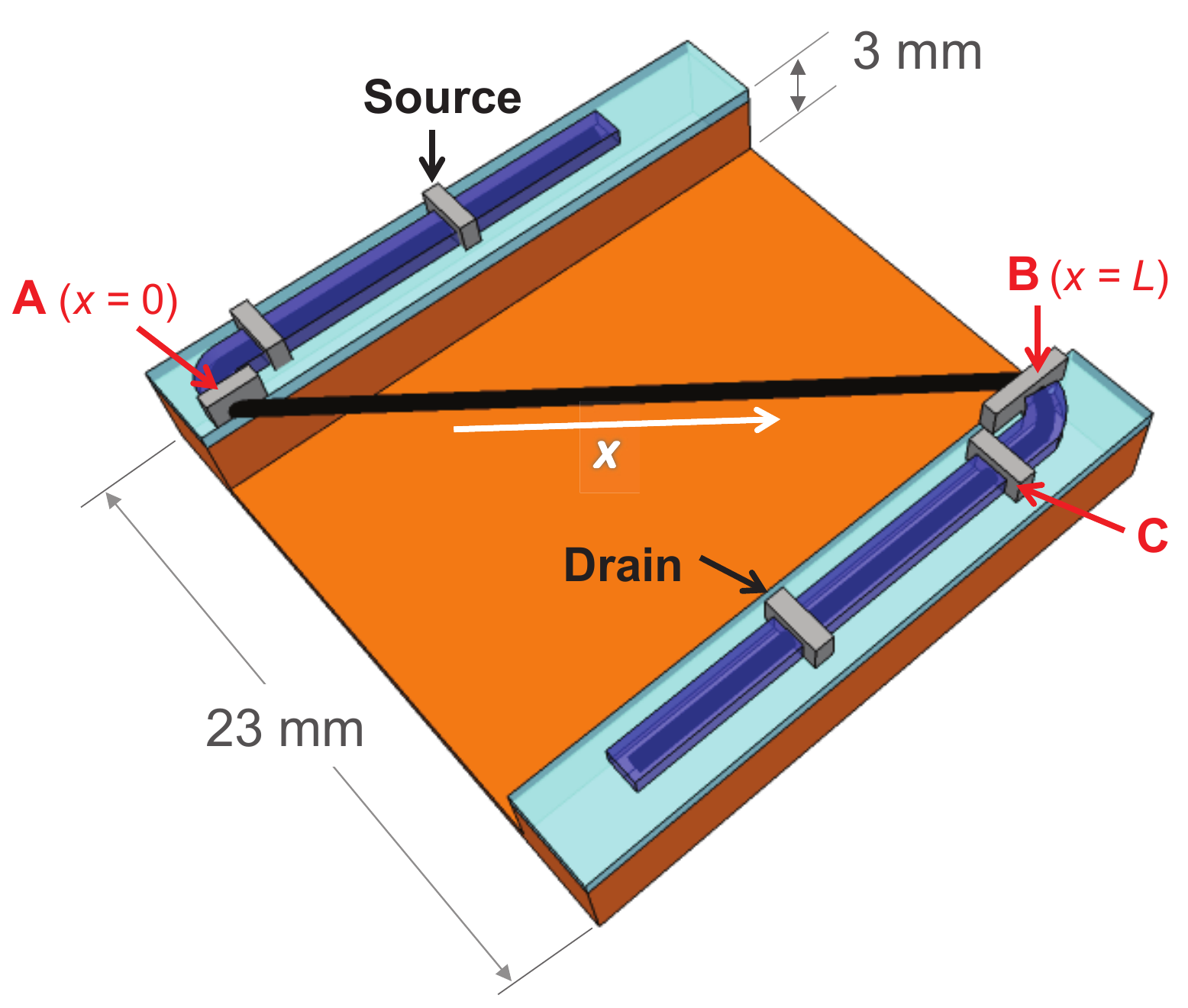}
\caption{Schematic diagram for the experimental setup used in this work.}
\label{fig4}
\end{center}
\end{figure}

A copper substrate with a wide river bed and two narrow river banks was used to hang the cable  (either a CNT fiber or copper wire)  as shown in Fig.~\ref{fig4}.  The depth of the bed was ${\sim}$3~mm and the width ${\sim}$23~mm.  The cable was bent into a Z-turn with its mid-section suspended over the river bed. Each arm was placed on a thin (100~$\mu$m) electrically-insulating quartz slide, which was itself placed on the river bank.  Several silver epoxy electrodes (${\sim}$0.5~mm wide) were placed on the fiber for resistivity measurements.  In particular, two electrodes (denoted by A and B in Fig.~\ref{fig4}) were placed at the ends of the suspended portion of the cable.  Electrode C  was placed 2~mm away from electrode A and B, and BC serves as a local probe to monitor the resistivity change at the end of the suspended fiber.  The whole device was assembled on a vacuum-sealed heating-cooling stage in which the temperature of the device and the gas environment could be adjusted.

\subsection{Current sweeping procedure}

\begin{figure}
\begin{center}
\includegraphics[scale=0.55]{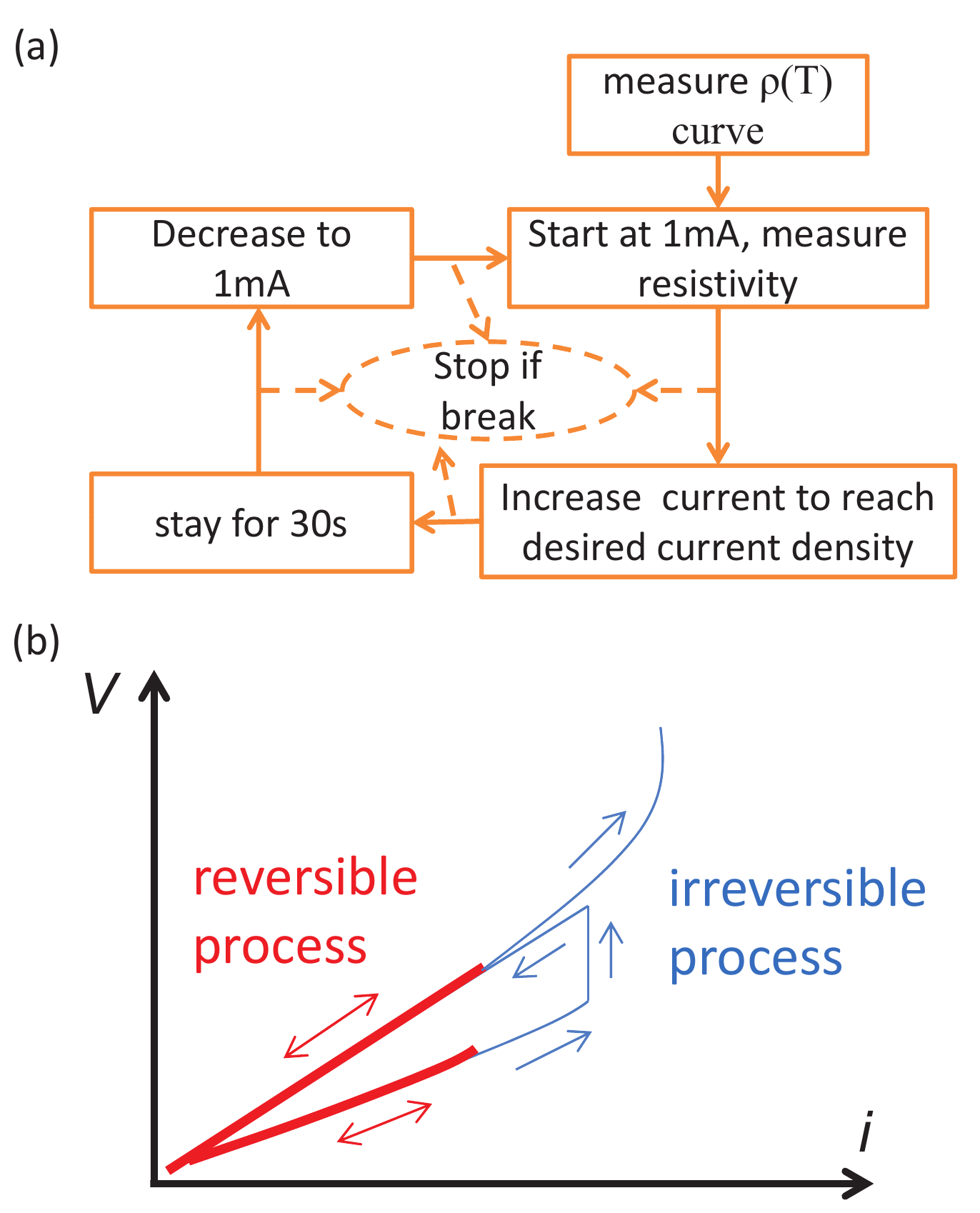}
\caption{Experimental procedure for studying high-field DC transport to determine the CCC of CNT fibers.  (a)~Flow chart describing the procedure step-by-step.  (b)~Schematic diagram showing reversible and irreversible processes in a current-sweeping cycle.}
\label{fig5}
\end{center}
\end{figure}

The experimental procedure is shown in the flow chart in Fig.~\ref{fig5}(a).  First, the sample is uniformly heated up and cooled down by the heating-cooling stage, while the resistivity of of the AB section is measured as a function of temperature [see Fig.~\ref{fig1}(a)].  Then a current sweep is carried out in high vacuum (10$^{-5}$ Torr), in one of four dry gases (nitrogen, helium, argon, and air) at atmospheric pressure, or on an intrinsic silicon substrate in air.  In this step, the current is gradually swept up to reach the desired current density, held for 30 seconds, and then gradually swept down to 1~mA [see \ref{fig5}(a)].  The experiment is halted if the fiber breaks.  If the electrical properties of the fiber are unchanged by current sweeping, the I-V curves from all sweeping cycles show no hysteresis (Fig.~\ref{fig5}(a), red curve), but if they are changed, the I-V curve will initially follow the sweeping-down curve of the previous circle and then start to deviate (Fig.~\ref{fig5}(b), blue curve).  Since the fiber might not be homogeneously Joule heated, the extracted resistivity is an average value given by
\begin{equation}
\label{eq_avrgT0}
\overline{\rho}_{f}(i) = \frac{V_{f}}{i l_{f}},
\end{equation}
where $i$, $V$, and $l$ are the current density, voltage, and length of the cable with $f$ = AB and BC representing the measured section (see Fig.~\ref{fig4}).

\section{RESULTS AND DISCUSSION}

\subsection{Determination of failure current density (FCD) in vacuum}

\begin{figure*}
\begin{center}
\includegraphics[scale=0.75]{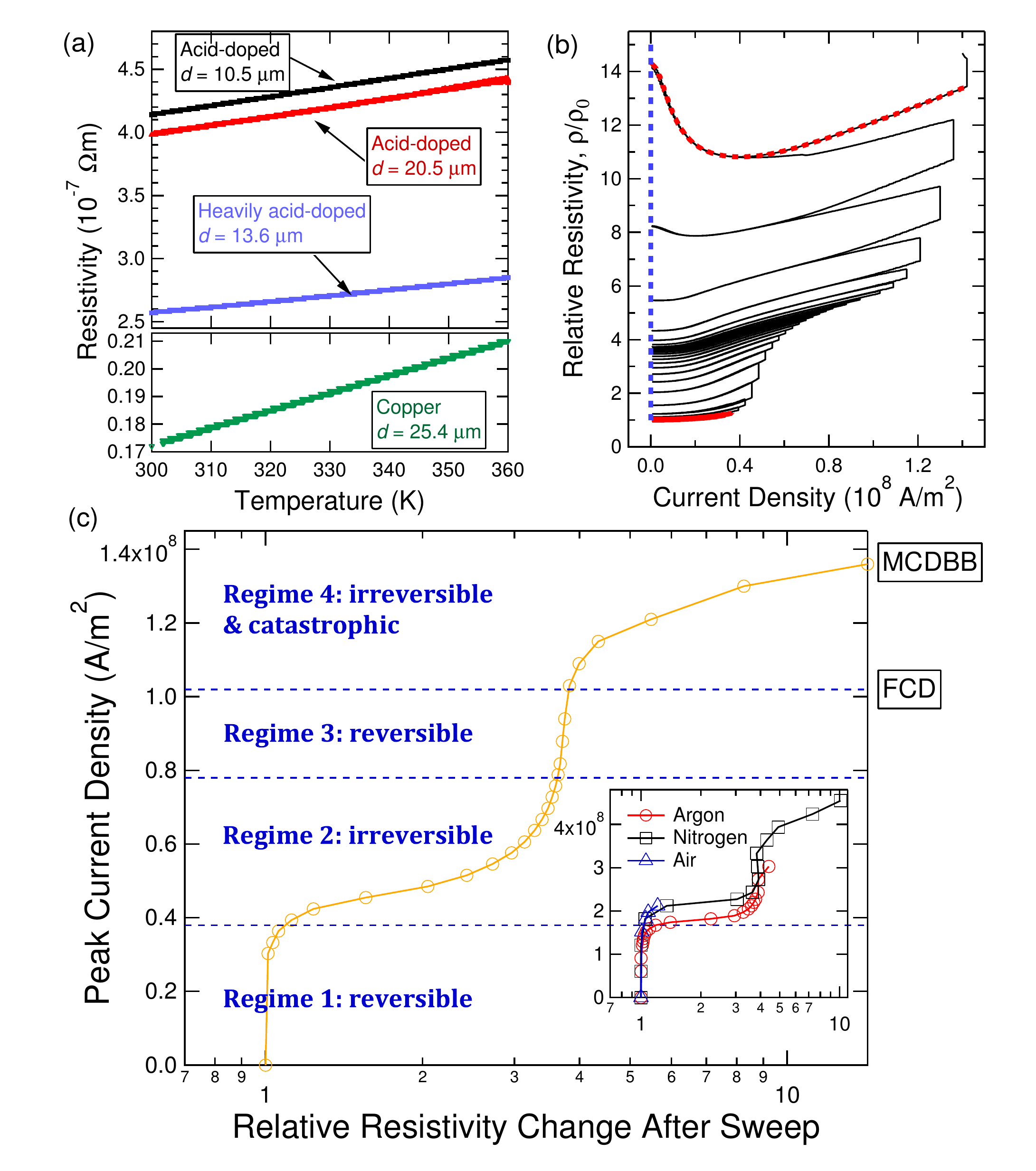}
\caption{(a)~Resistivity versus temperature from room temperature to 360~K for the highly conductive carbon nanotubes fibers studied.  The lowest curve is for the reference copper wire with a diameter of 25.4~$\mu$m.  (b)~Resistivity versus current density for the 20-$\mu$m diameter, acid-doped CNT fiber in vacuum, showing hysteresis loops through a number of current-sweeping cycles described in the Methods section, indicating irreversible changes induced by the large currents.  $\rho_0$ = 3.98 $\times$ 10$^{-7}$~$\Omega$m is the initial resistivity.  The red curve indicates the first sweep point.  The red dashed curve represents a typical curve in Regime 4.  (c)~Peak current density applied in a current-sweeping cycle versus the low-current resistivity measured after the cycle.  This graph shows four distinct regimes.  The definitions of the maximum current density before breaking (MCDBB) and failure current density (FCD) are also indicated.  The main panel is for data taken vacuum and the inset shows data taken in argon, nitrogen and air.}
\label{fig1}
\end{center}
\end{figure*}

\begin{figure*}
\begin{center}
\includegraphics[scale=0.85]{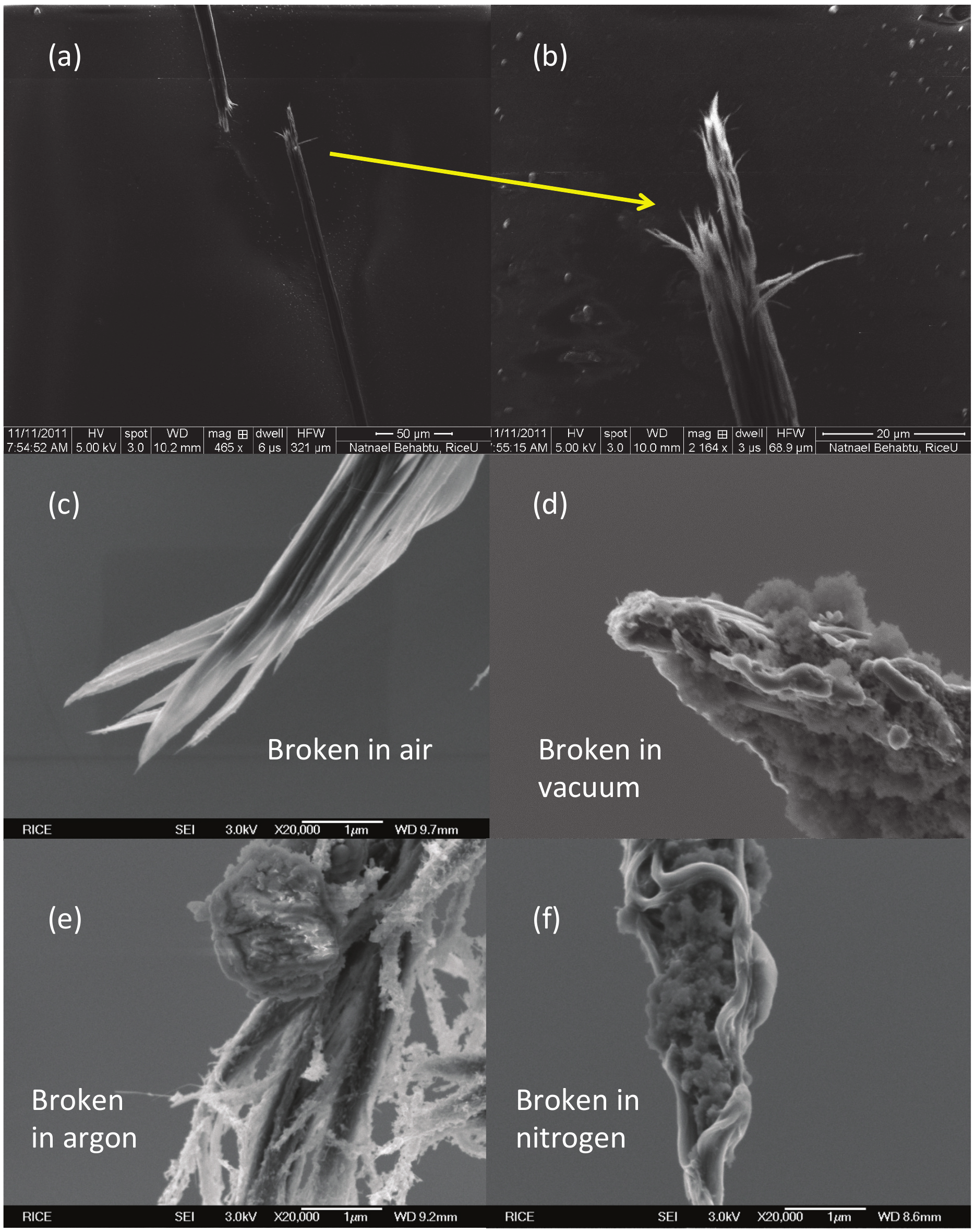}
\caption{(a) and (b): Scanning electron microscope images of a typical carbon nanotube fiber broken into two pieces in the middle due to high-current-induced Joule heating.  Magnified images of the breaking point of fibers that were broken in (c)~air, (d)~vacuum, (e)~argon, and (f)~nitrogen.}
\label{fig2}
\end{center}
\end{figure*}

Figure \ref{fig1}(a) shows the resistivity versus temperature for the three highly-conductive CNT fibers\cite{BehabtuetAl13Science} we studied, together with that for the reference copper wire with a diameter of 25.4~$\mu$m.   The fibers we tested were electrically p-doped by the presence of sulfur and chlorine inside the fiber, which is a residuum from the chlorosulfonic acid solvent used in the fabrication of the fibers.  The heavily acid-doped fiber contained more acid, making them less resistive.  The room temperature resistivity for the heavily acid-doped fiber was about 2.57~$\times$~10$^{-7}$~$\Omega$m while that of the 10.5~$\mu$m diameter (20.5~$\mu$m diameter) acid-doped fiber was 4.12 (3.98) $\times$ 10$^{-7}$~$\Omega$m.  On the other hand, the mass density of the heavily-doped sample was 1.5~$\times$~10$^{3}$~kg/m$^3$, as compared to 1.2~$\times$~10$^{3}$~kg/m$^3$ of the other acid-doped fibers.   Another issue with fiber doping is its stability. We notice that in several cases, excessive acid doping makes the room temperature resistivity of several fibers as low as 1.7~$\times$~10$^{-7}$~$\Omega$m, but at the same time, after annealing at 373~K, this value quickly returned back to about 2.5~$\times$~10$^{-7}$~$\Omega$m.  Thus, the temperature range in which a fiber can be operated without any irreversible property change must be considered.   The room temperature resistivity of copper is approximately one order of magnitude less than that of fibers.  Its accepted value is 1.725~$\times$~10$^{-8}$~$\Omega$m while the measured value here was 1.74~$\times$~10$^{-8}$~$\Omega$m.  In all cases, the resistivity ($\rho$) linearly increased with temperature ($T$), i.e.,
\begin{equation}
\label{eq_linear}
\rho(T) = \rho(T_{\rm amb}) ( 1 + \alpha \cdot \Delta T ),
\end{equation}
where $T_{\rm amb}$ is the ambient temperature, $\Delta T \equiv T - T_{\rm amb}$ is the temperature measured from $T_{\rm amb}$, and $\alpha$ is a positive constant.  This equation provides us with a convenient means for monitoring the temperature rise as a result of current-induced heating.

Figure \ref{fig1}(b) shows the resistivity as a function of current density for the 20-$\mu$m-diameter acid-doped fiber in vacuum for $T_{\rm amb}$ = 303~K.  The resistivity is normalized to the initial value, $\rho_0$ = 3.98 $\times$ 10$^{-7}$~$\Omega$m, before the fiber is heated.  As the current density increases, the temperature increases through Joule heating, which in turn increases the resistivity through Fig.~\ref{eq_linear}.  Starting from the lowest red curve, after a number of current-sweeping cycles, the resistivity versus current density curve undergoes irreversible changes with uneven paces.  To better visualize this process, Fig.~\ref{fig1}(c) plots the highest applied current density against the resistivity measured at a very low current density after each sweeping cycle [along the blue dashed line in Fig.~\ref{fig1}(b)].

Interestingly, Fig.~\ref{fig1}(c) reveals four distinct regimes as the current density is gradually increased from zero toward the ultimate value at which the fiber eventually breaks.  In Regime 1, the I-V curve is reversible, and thus, the resistivity does not change after each sweeping cycle.  In Regime 2, a drastic irreversible process takes place, and the resistivity permanently increases by about 4 times.  In Regime 3, the {\em I-V curve becomes reversible again}, showing stable properties of a new, current-annealed fiber; i.e., the acid, which is an effective dopant, is removed by heating.  Finally, in Regime 4, the resistivity starts increasing very rapidly until the fiber breaks.  Given sufficient time, any current-density value in Regime 4 ultimately leads to fiber breaking.

We define the current-density value that corresponds to the boundary between Regimes 3 and 4 as the failure current density, or FCD, of the fiber, which is different from the maximum current density before breaking (MCDBB).  See Fig.~\ref{fig1}(c) for the difference between FCD and MCDBB.  The value of MCDBB is ill-defined and can have a large uncertainty, depending on such experimental details as the sweeping speed, step size, sweeping method (current or voltage), and sweeping pattern.  One the other hand, FCD  is determined by the characteristic temperature limits beyond which the quality of the fiber is altered.  These temperature limits can be considered intrinsic because they reflect such fiber properties as the defect density, impurities, alignment, etc.  Therefore, FCD is a better quantity for characterizing the CCC of wires.  Unfortunately, all previously reported CCC for aligned buckypapers,\cite{ParketAl08Nanotech} carbon fibers,\cite{ParketAl08Nanotech} and CNT fibers\cite{ParketAl08Nanotech,SongetAl12Nanotechnology} are MCDBB values.  The FCD value for the particular case shown in Fig.~\ref{fig1}(c) is 1.03 $\times$ 10$^8$~A/m$^2$, while the MCDBB is $\sim$1.4 $\times$ 10$^8$~A/m$^2$.

\subsection{Joule heating induced annealing and irreversible change in I-V curves}

The data in Fig.~\ref{fig1}(c) provides significant insight into the mechanism by which the fiber leads to a catastrophic failure at high current densities.  In Regime 2, a drastic irreversible process occurs, and the resistivity becomes about four times larger than the original value.  Based on our modeling (see Appendix for detail), the maximum temperature of the fiber at the boundary between Regimes 1 and 2 is about 470~K, which is higher than the boiling point of chlorosulfonic acid (423~K).  Therefore, we attribute the increase of resistivity in Regime 2 to an irreversible reduction of charge carriers through removal of chlorosulfonic acid, a p-type dopant.  Even though Regime 2 is fairly narrow, the increase of resistivity significantly increases Joule heating.  As a result, the finishing temperature (at the boundary between Regimes 2 and 3) is estimated to be $\sim$1305~K.  

Notably, the quality of the fiber is not degraded during the heating process in Regime 2, as confirmed by the maintained G/D ratio in Raman spectra (shown in Appendix).  The fact that the I-V curve is reversible again in Regime 3 indicates that no chemical changes happen in this regime and the fiber is stable even though it is heated by high currents.  Thus, we are essentially dealing with an annealed fiber in this regime, whose properties are different from those of the original, acid-doped fiber in Regime 1.  Finally, as the average temperature exceeds $\sim$1740~K, the fiber enters Regime 4, and the resistivity increases very rapidly, until it breaks.

A close examination of $\rho$--$i^2$ curves in Regime 4 provides further insight into the final moments when the fiber is breaking apart.  Note that a typical curve in Regime 4, e.g., the red dashed curve in Fig.~\ref{fig1}(b), shows a qualitatively different trend than those in the other three regimes; that is, the {\em resistivity initially decreases and then increases with increasing current density}.  This unusual trend can be explained only if we consider the general temperature dependence of resistivity of these fibers in a wider temperature range.  Namely, with increasing temperature (from, e.g., 4.2~K), the resistivity initially decreases due to thermally-driven hopping transport and then increases due to intra-tube phonon-carrier scattering.\cite{BehabtuetAl13Science}  The crossover temperature ($T^*$), where the resistivity is minimal, is lower than the ambient temperature in the fibers we study here.  This is why Fig.~\ref{eq_linear} holds in the 300-360~K range.  However, when the fiber starts breaking, the $\rho$--$T$ curve itself starts changing irreversibly and dynamically.  Specifically, as the fiber starts structurally deteriorating, the hopping transport contribution becomes more and more important in determining the resistivity, which pushes $T^*$ higher and higher during the breaking process.  The fact that we see an initial decrease in resistivity in Regime 4 (e.g., the green curve) is evidence that $T^*$ has already become higher than $T_{\rm amb}$.  This is a self-intensified process because the initial damage forces a higher current through the remaining conductive paths, accelerating the breaking process.

\subsection{Dependence of failure current on the type of surrounding gas}

The inset of Fig.~\ref{fig1}(c) shows that the type of gas surrounding the fiber critically affects the boundaries of different current regimes.  In an argon or nitrogen gas environment, our CNT fibers exhibit qualitatively the same behavior as in vacuum, but the boundaries of regimes are shifted to much higher values.  This is understandable because the gas contributes convective cooling, whereas in vacuum black-body radiation is essentially the only thermal path, except through the end contacts.  On the other hand, fibers break more easily in air, usually breaking already in Regime 2; i.e., the FCD value corresponds to the boundary between Regimes 1 and 2.  We attribute this reduced FCD in air to the oxidation of carbon nanotubes, which can happen at temperatures between 773 and 873~K.  The FCD values measured in different environments are summarized in Table~\ref{table_FCD}.

\begin{table}
  \caption{Failure current density (FCD), maximum current density before breaking (MCDBB), and specific failure current (SFC) values determined for an acid-doped carbon nanotube fiber with a diameter of 20.5~$\mu$m, through measurements in vacuum, argon, nitrogen, and air.  The values in parentheses are the corresponding values estimated for copper (see Appendix for details on the estimations).}
  \label{table_FCD}
  \begin{tabular}{l|cccc}
    \hline
    \hline
     & Vacuum & Argon & Nitrogen & Air\\
     \hline
     FCD (GA/m$^2$) & 0.103 & 0.273 & 0.330 & 0.182\\
     & & (0.912) & (1.135) & (1.135) \\
     \hline
     MCDBB (GA/m$^2$) & 0.136 & 0.303 & 0.455 & 0.211\\
     \hline
     SFC (A$\cdot$m/g) & 85.0 & 225 & 272 & 150 \\
     & & (66.3) & (82.5) & (82.5) \\
    \hline
    \hline
  \end{tabular}
\end{table}

It can be seen that the FCD values of the copper wire are still higher than those for the CNT fibers,  mainly due to the lower resistivity of copper.  However, the SFC values of the CNT fibers exceed copper wires' in all tested gases due to the much smaller mass densities of the CNT fibers.  The dependence of FCD on the fiber dimensions as well as the surrounding medium are properly explained through our theoretical analyses (see Appendix B and Appendix C for detail; briefly, the FCD is proportional to $d^{-0.913}$ for a gas medium and $d^{-0.5}$ for a solid medium).  These analyses also enable us to make a comprehensive comparison of CCC, mainly the MCDBB, with those numbers reported for other macroscopic CNT devices (see Appendix E).  The MCDBB value of this fiber is much higher than that of aligned buckypaper.\cite{ParketAl08Nanotech}  It is also higher than those reported for both a carbon fiber and CNT fiber\cite{SongetAl12Nanotechnology} while it is similar to another reported value for a CNT fibril\cite{ZhaoetAl11SciRep} (although the length of the sample was not specified).  It is much smaller than a recently reported value 6~$\times$~10$^{12}$~A/m$^2$ for a CNT-Cu composite, in which CNT occupies 45\% volume fraction.\cite{SubramaniametAl13NC}

The scanning electron microscopy images in Fig.~\ref{fig2} are further evidence that the type of surrounding gas affects the way the fiber breaks, also providing some clues about the breaking mechanisms.  The break usually occurs in the middle of the fiber, where the local temperature is the highest, consistent with our simulations described in the next section [see Fig.~\ref{fig3}(a)].  Some sharp, needle-like tips are seen at the breaking point, similar to earlier reports.\cite{WeietAl07NL,SongetAl12Nanotechnology,DingetAl07NL}  
Tips observed in the fibers broken in vacuum, nitrogen, and argon are decorated by some ball-like clusters, whereas tips formed in the fiber broken in air are observed to be clean and free from such clusters.   We hypothesize that these clusters are amorphous carbon or other carbon structures produced under high temperature.  When the fiber breaks in air, the breaking mechanism is likely through oxidation, the product is CO$_2$, so there is no amorphous carbon residual.   However, in the other three cases (vacuum, argon, and nitrogen), the fiber breaks from the inside through current-induced heating that raises the temperature to the sublimation point.  Thus, the generated amorphous carbon cannot escape to the outside quickly and remains  at the breaking point.   


\subsection{Thermal conductance between the fiber and its surrounding gas}

Since current-induced heating is the main cause of resistivity change in the fibers under high current densities, it is the balance between the heating current and the thermal dissipation into the surroundings that determines the detailed behavior of nonlinear I-V curves.   We have developed a model that can make quantitative predictions and yield important parameters through fitting experimental data.  Here we determine the thermal conductance, $g$, between the fiber and the surrounding gas through analysis of the nonlinear I-V in the reversible range (Regime 1) for the heavily acid-doped fiber in argon, nitrogen, and helium.

\begin{figure}
\begin{center}
\includegraphics[scale=0.6]{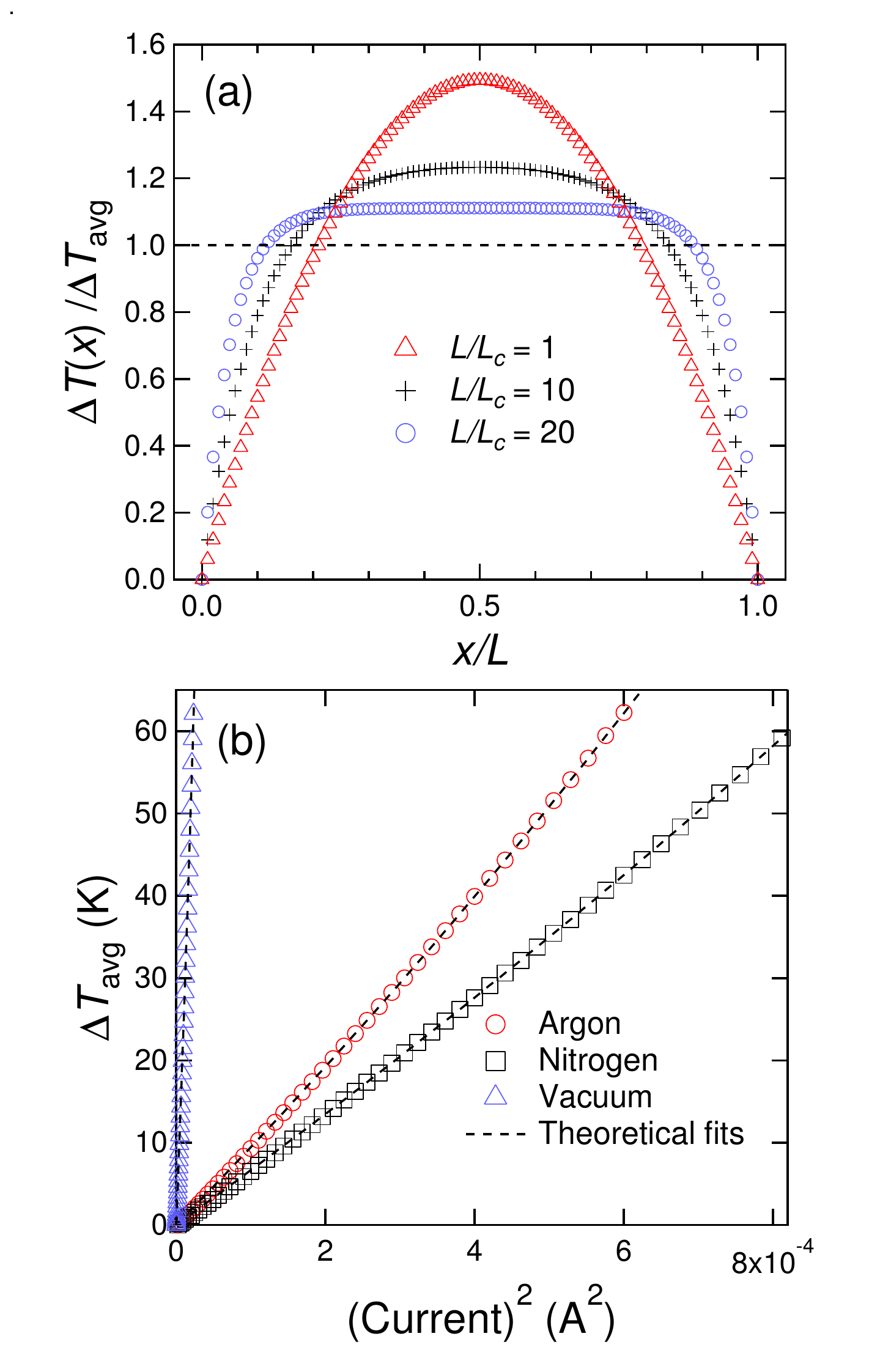}
\caption{(a)~Temperature distribution in a current-carrying wire with finite length ($L$) with different critical thermal lengths ($L_{\rm c}$), calculated based on a model described in Appendix.  (b)~Measured average temperature increase of the heavily acid-doped CNT fiber as a function of the square of the applied current in vacuum, nitrogen, and argon.  The dashed lines are theoretical fits based on Eq.~(\ref{eq_avrgT}).}
\label{fig3}
\end{center}
\end{figure}

\begin{table*}
  \caption{Values for thermal conductance ($g$) between the current-carrying fiber and surrounding gas for the cases of helium, nitrogen, and argon, determined for carbon nanotube fibers and copper cable.  $d$: fiber diameter.  $\kappa$: thermal conductivity.  The last two rows show estimates for CNT fibers with $d$ = 10.5 and 20.5 $\mu$m, respectively, from our model based on natural convection theory (see Appendix for detail).}
  \label{table_conductance}
  \begin{tabular}{lccccc}
    \hline
    \hline
     & $d$ & $\kappa$ & $g_{\rm helium}$ & $g_{\rm nitrogen}$ & $g_{\rm argon}$ \\
     & ($\mu$m) & (W/mK) & (W/m$^2$K) & (W/m$^2$K) & (W/m$^2$K) \\
     \hline
     Heavily acid-doped CNT fiber & 13.6 & 420 & 2973 & 571 & 395 \\
     Acid-doped CNT fiber & 10.5 & 331 & 3689 & 716 & 494 \\
     Acid-doped CNT fiber & 20.5 & 314 & 2007 & 379 & 260 \\
     Copper wire & 25.4 & 423 & 2024 & 397 & 260 \\
     Theoretical & 10.5 & & 3928 & 879 & 606 \\
     Theoretical & 20.5 & & 2260 & 506 & 349 \\
     \hline
     \hline
  \end{tabular}
\end{table*}

A current-carrying wire with finite length has a temperature distribution along the current direction, with the highest temperature in the middle of the wire and the lowest temperature at both ends [Fig.~\ref{fig3}(a)].  The average temperature increase relative to the ambient temperature, ${\Delta T}_{\rm avg}$ = $T_{\rm avg} - T_{\rm amb}$, of a current-carrying wire with length $L$ at a current density $i$ is given by
\begin{equation}
\label{eq_avrgT}
{\Delta T}_{\rm avg} = {\Delta T}_{\infty} \left \{1 - \frac{\tanh(L/2L_{\rm c})}{L/2L_{\rm c}} \right \} + \Delta T_{\rm end}\frac{\tanh(L/2L_{\rm c})}{L/2L_{\rm c}},
\end{equation}
\begin{equation}
\label{eq_infty}
{\Delta T}_{\infty} = {i^2 \rho(T_{\rm amb}) L^2_{\rm c} \over \kappa},
\end{equation}
\begin{equation}
\label{eq_critical_length}
L_{\rm c} \equiv \sqrt{\frac{\kappa d}{4g -i^2 \rho(T_{\rm{amb}}) d \alpha} },
\end{equation}
where $T_{\rm amb}$ = 300~K is the ambient temperature, ${\Delta T}_{\infty}$ is the temperature increase when the length is infinite ($L \rightarrow \infty$), $L_{\rm c}$ is the critical thermal length, $\kappa$ is the thermal conductivity of the wire at the ambient temperature (whose temperature dependence is very small in this temperature range), $\alpha$ is the temperature coefficient of $\rho(T)$ appearing in Eq.~(\ref{eq_linear}), $d$ is the wire diameter, and $\Delta T_{\rm end}$ is the temperature increase at the end contacts induced by the current.  
See Appendix B for details.

In our experiments, we obtain both ${\Delta T}_{\rm avg}$ and $\Delta T_{\rm end}$ as a function of $i^2$ by converting the resistivities of the `AB' and `BC' portions, respectively, of the device (see Fig.~\ref{fig4}) into temperatures {\it via} Eq.~(\ref{eq_linear}).  Then, by fitting such data with Eq.~(\ref{eq_avrgT}), we determine $\kappa$ and $g$.  Particularly, under vacuum at a pressure of 10$^{-5}$~Torr, heat transfer between the wire and gas can be ignored ($g=0$), in which case the only fitting parameter is $\kappa$.  Figure~\ref{fig3}(b) shows fitting results for ${\Delta T}_{\rm avg}$ obtained for the heavily acid-doped CNT fiber in vacuum, argon, and nitrogen.  The parameters extracted are listed in Table~\ref{table_conductance}; the extracted $\kappa$ value is in good agreement with that reported earlier.\cite{BehabtuetAl13Science}  

Listed in the last two rows in Table~\ref{table_conductance} are estimates for the $g$ values for the 10.5-$\mu$m and 20.5-$\mu$m diameter CNT fibers based on natural convection theory (see Appendix), providing the correct orders of magnitude as well as the observed trends for $g$ with respect to the fiber diameter and gas type.  In particular, we find that $g$ increases with decreasing $d$ ($g_{\rm gas} \propto d^{-0.826}$), which in turn makes the CCC of smaller-diameter fibers larger, in agreement with the experimental data.  Our results also show that CNT fibers possess similar $g$ values to copper wires, implying that the inner structure of the fibers is irrelevant to $g$, i.e., any minimal gas flow into the body of the CNT fiber does not contribute to heat transfer.  This fact supports the notion that our fibers consist of tightly-packed and well-aligned CNTs.   

\subsection{Estimation of the continuous current rating (CCR)}

\begin{table*}
  \caption{Continuous current rating, $I^{\rm{CCR}}$ (mA), and the corresponding specific continuous current rating, $J^{\rm{CCR}} = I^{\rm{CCR}} / \lambda$ (A$\cdot$m/g), estimated for the CNT fibers and copper wire for an operating temperature of 363~K.  The diameter $d$ (in $\mu$m) and linear mass density $\lambda$ (in 10$^{-3}$\,g/m) are also listed for each fiber.}
  \label{table_CCR}
  \begin{tabular}{l||cc|ccc|ccc}
     & $d$ & $\lambda$ & $I^{\rm CCR}_{\rm helium}$ & $I^{\rm CCR}_{\rm nitrogen}$ & $I^{\rm CCR}_{\rm argon}$ & $J^{\rm CCR}_{\rm helium}$ & $J^{\rm CCR}_{\rm nitrogen}$ & $J^{\rm CCR}_{\rm argon}$ \\
     \hline
     Heavily acid-doped fiber & 13.6 & 0.212 & 62.2 & 27.3 & 22.7 & 293 & 129 & 107 \\
     Acid-doped fiber & 10.5 & 0.105 & 37.3 & 16.4 & 13.6 & 355 & 156 & 130 \\
     Acid-doped fiber & 20.5 & 0.400 & 76.5 & 33.2 & 27.5 & 191 & 83.1 & 68.8 \\
     Copper wire & 25.4 & 4.54 & 482 & 214 & 173 & 106 & 47.0 & 38.1
  \end{tabular}
\end{table*}

With the above determined values for $g$, together with the temperature-dependent resistivity, Eq.~(\ref{eq_linear}), we can now calculate the CCR values for the CNT fibers for any assumed operating temperature in the reversible current region, $T_{\rm o}$.   As an example, here we show the calculated CCR for infinitely long CNT fibers at $T_{\rm o}$ = 363~K.  By setting $\Delta T_{\rm \infty} = T_{\rm o} - T_{\rm amb}$ in Eq.~(\ref{eq_infty}), we can derive the following compact expression for calculating CCR:
\begin{equation}
\label{eq_CCC1}
  I^{\rm{CCR}} = \sqrt{ { \pi^2 d^3 (T_{\rm o} - T_{\rm amb}) g} \over {4 \, \rho (T_{\rm o})}}.
\end{equation}
Note that with $g \propto d^{-0.826}$ we can predict that $I^{\rm CCR} \propto d^{1.087}$.  The specific CCR is then given as $J^{\mathrm{CCR}}=I^{\mathrm{CCR}}/\lambda$, where $\lambda$ is the linear mass density (in kg/m) of the cable.  Table~\ref{table_CCR} lists the obtained $I^{\mathrm{CCR}}$ and $J^{\mathrm{CCR}}$ values for our CNT fibers and the reference copper wire for $T_{\rm o}$ = 363~K and $T_{\rm{amb}}$ = 303~K.  Again, the specific CCR values of CNT fibers are comparable to or larger than that of the copper wire.  The current densities corresponding to these CCR values are 2--5 $\times$ 10$^8$~A/m$^2$ for the 10-$\mu$m fiber and 1--2 $\times$ 10$^8$~A/m$^2$ for the 20-$\mu$m fiber.

\section{CONCLUSION}

We characterized the current-carrying capacity, CCC or ampacity, of highly-conductive carbon nanotube fibers by measuring both their failure current density, FCD, and continuous current rating, CCR, based on a systematic study of their DC transport performance over a broad current range.  We first demonstrated that there are four distinct regimes as we gradually increase the current density toward the breaking point; we provided qualitative explanations for the different regimes.  The measured FCD values ranged from 10$^7$ to 10$^9$~A/m$^2$, depending on the diameter as well as the environmental gas type.  The measured FCD in vacuum was the lowest due to lack of heat exchange with gases, while the measured FCD in air was smaller than in the other tested gases because of oxidation.  In addition, we theoretically analyzed the heat exchange between CNT fibers and each type of gas, successfully extracting the thermal conductance between them.  Furthermore, we proved that the tight packing and good alignment of our CNT fibers are crucial for obtaining the large CCC values.   The FCD of our fibers is higher than previously reported carbon fibers and CNT fibers.   If normalized by the linear mass density, both the specific failure current and specific continuous current rating of these CNT fibers exceed those of the copper wire.   The combination of high specific CCC and good mechanical strength of our CNT fibers makes them promising candidates for power transmission cables.

\section*{Acknowledgements}

This work was supported by the Department of Energy (through Grant No.~DE-FG02-06ER46308), the National Science Foundation (through Grants No.~OISE-0530220 and EEC-0540832), the Robert A. Welch Foundation (through Grant Nos.~C-1509 and C-1668), Teijin Aramid BV, Teijin Techno Products Ltd., Air Force Office of Scientific Research FA9550-09-1-0590, and the U.S. Department of Defense National Defense Science and Engineering Graduate Fellowship.


\appendix

\section{Modeling the temperature distribution in a current-carrying fiber}

The temperature distribution of a suspended current-carrying fiber can be simulated through a one-dimensional energy balance equation written for the temperature of the fiber as a function of position ($x$) and time ($t$), $T(x,t)$:
\begin{widetext}
\begin{equation}
\label{eq_heat}
\frac{\kappa \partial^{2} T(x,t)}{\partial x^{2}}+i(t)^2 \rho \left(T(x,t) \right)-\frac{4 g \cdot (T(x,t)-T_{\rm amb})}{d}-\frac{4 \sigma \epsilon \cdot (T^4(x,t) -T^4_{\rm amb})}{d} = C_{\rm v}\frac{ \partial  T(x,t)}{\partial t}
\end{equation}
\end{widetext}
with the initial condition $T(x,0) = T_{\rm amb}$,  where $\kappa$ is the thermal conductivity, $i$ is the current density, $\rho$ is the resistivity, $g$ is the thermal conductance between the fiber and its surrounding, $d$ is the fiber diameter, $T_{\rm amb}$ is the ambient temperature, $\sigma$ is the Stefan-Boltzmann constant, $\epsilon$ is the emissivity, and $C_{\rm v}$ is the volume heat capacity.  The four terms on the left-hand side of Eq.~(\ref{eq_heat}) represent heat diffusion, Joule heating, thermal exchange with the surrounding, and blackbody radiation, respectively. In all four regimes defined in Fig.~\ref{fig1}(c), heat exchange through blackbody radiation is negligible compared to heat exchange with the surrounding; it is also negligible in vacuum below 360~K compared to heat exchange through the end contacts.

Before current is applied, the whole system is under thermal equilibrium at $T(x,0) = T_{\rm amb}$.  As the current, $i(t)$, increases as a function of time, it also increases the temperature, $T(x,t)$, as a function of time.  However, we are more interested in the situation when the current value is maintained for a certain period of time for each sweeping-current step, i.e., $\frac{ \partial i(t)}{\partial t} = 0$.  If the system can reach a new thermal equilibrium afterwards, then Eq.~(\ref{eq_heat}) becomes a stationary heat equation with $\frac{ \partial T(x,t)}{\partial t}=0$. This is what happens in Regimes 1 and 3, where $T(x)$ can be written as a function of $i$; given any restriction on $T(x)$, such as the operational temperature, the corresponding $i$ can be determined.  On the other hand, in Regimes 2 and 4, no thermal equilibrium can be reached, and $T(x,t)$ depends not only on $i$ but also on $t$.  In particular, for any current value in Regime 4, $T(x,t)$ will keep increasing until the fiber breaks. This makes the maximum current density before breaking (MCDBB) ill-defined, since it depends on the details of sweeping conditions.

\section{Determining the thermal conductance, $g$, between the fiber and its surrounding gas 
}

We consider a current-carrying fiber suspended over a region of $x$ = 0 to $L$ (see Fig.~\ref{fig4}) with the current density value corresponding to Regime 1 (see Fig.~\ref{fig1}c). Under the stationary condition, $\frac{ \partial T(x,t)}{\partial t}=0$, Eq.~(\ref{eq_heat}) reduces to
\begin{equation}
\label{eq_heat1}
\frac{\kappa \partial^{2} \Delta T(x)}{\partial x^{2}}+i^2 \rho \big (T(x)\big )-\frac{4 g \Delta T(x)}{d}
= C_{\rm v}\frac{ \partial \Delta T(x)}{\partial t}
\end{equation}
where $\Delta T(x) \equiv T(x)-T_{\rm amb}$ is the current-induced temperature rise as a function of position, measured relative to the ambient temperature (= 303~K).  Using the boundary conditions of $\Delta  T_{\rm{end}}=\Delta T(0)=\Delta T(L)$, one can solve Eq.~(\ref{eq_heat1}) to obtain~\cite{ShietAl09JAP}
\begin{widetext}
\begin{equation}
\label{eq_temp}
  \Delta T(x) = \left [ 1 - \frac{\cosh \{(x-L/2)/L_{\rm c}\}}{\cosh(L/2L_{\rm c})} \right ] \Delta T_{\infty}+\Delta T_{\rm{end}} \frac{\cosh \{(x-L/2)/L_{\rm c}\}}{\cosh(L/2L_{\rm c})},
\end{equation}
\end{widetext}
where 
\begin{equation}
\label{eq_infty}
{\Delta T}_{\infty} = {i^2 \rho(T_{\rm amb}) L^2_{\rm c} \over \kappa},
\end{equation}
is the temperature increase when the length is infinite ($L \rightarrow \infty$) and
\begin{equation}
\label{eq_critical_length}
L_{\rm c} \equiv \sqrt{\frac{\kappa d}{4g -i^2 \rho(T_{\rm{amb}}) d \alpha} },
\end{equation}
is the critical thermal length.  $\Delta T_{\rm end}$ can be estimated through the temperature change of the fiber between the two contacts `B' and `C' (see Fig.~\ref{fig4}).

The measured resistivity is the average value over the length, $L$, of the fiber section suspended between the two terminals `A' and `B' (see Fig~.\ref{fig4}):
\begin{equation}
\label{eq_avrgT1}
\rho_{\rm avg} \equiv \frac{1}{L}\int_{0}^{L}\rho(T(x)) d x.
\end{equation}
Here, the temperature dependence of the resistivity
\begin{equation}\label{eq_linear}
\rho(T) = \rho(T_{\rm amb}) ( 1 + \alpha \cdot \Delta T ),
\end{equation}
is shown in Figure 1a of the article.  By substituting Eq.~(\ref{eq_linear}) into Eq.~(\ref{eq_avrgT1}), we obtain
\begin{widetext}
\begin{equation}
\label{eq_therm_co}
 \rho_{\rm avg}
=\frac{1}{L}\int_{0}^{L} \rho(T_{\rm{amb}}) \{ 1+\alpha \cdot
(T(x)-T_{\rm{amb}}) \}dx=\rho(T_{\rm{amb}})(1+\alpha \cdot \Delta T_{\rm avg}),
\end{equation}
\end{widetext}
where $\Delta T_{\rm avg}$ is defined as the average temperature change of the cable relative to $T_{\rm amb}$:
\begin{equation}
\label{AverageTemp}
\Delta T_{\rm avg} \equiv \frac{1}{L}\int_{0}^{L} T(x) dx-T_{\rm amb}.
\end{equation}
By combining Eqs.~(\ref{eq_temp}) and (\ref{AverageTemp}), we obtain
\begin{equation}\label{eq_avrgT}
{\Delta T}_{\rm avg} = {\Delta T}_{\infty} \left \{1 - \frac{\tanh(L/2L_{\rm c})}{L/2L_{\rm c}} \right \} + \Delta T_{\rm end}\frac{\tanh(L/2L_{\rm c})}{L/2L_{\rm c}}.
\end{equation}
$\Delta T_{\rm end}(i)$ is obtained through the current-density dependence of the resistivity of the `BC' section of the fiber (Fig.~\ref{fig4}), together with Eq.~(\ref{eq_linear}).  $\rho_{\rm avg}(i)$ is the current-density dependence of the resistivity of the `AB' section of the fiber (Fig.~\ref{fig4}), which can be converted into $T_{\rm avg}(i)$ through Eq.~(\ref{eq_therm_co}).  Finally, we use Eq.~(\ref{eq_avrgT}) to fit the $T_{\rm avg}(i)$ curve to determine $L_{\rm c}$, and hence, $\kappa$ and $g$ [see Eq.~(\ref{eq_critical_length})], as detailed below.  Note that here $\kappa$ and $g$ are assumed to be independent of $T$ because the overall temperature change is small; the characteristic temperature for $g$ should be the ``film temperature,'' Eq.~(\ref{eq_filmT}), which only varies by 30~K (see the next Appendix for more detail).

\begin{figure}
\begin{center}
\includegraphics[scale=0.57]{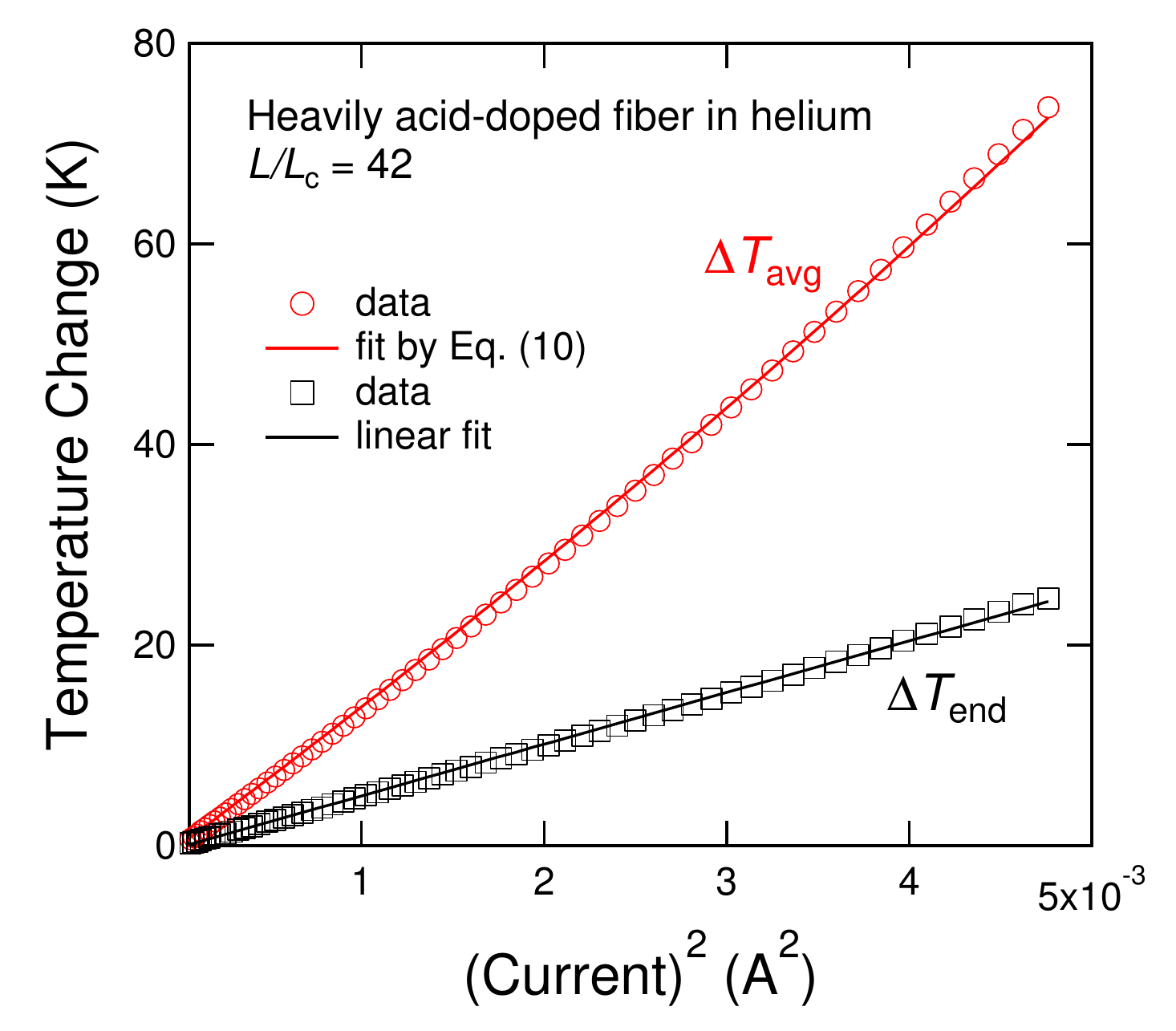}
\caption{Measured average temperature increase of the heavily acid-doped CNT fiber as a function of the square of the applied current in helium.  The dashed lines are theoretical fits.}
\label{figS1}
\end{center}
\end{figure}

We first fit $\Delta T_{\rm avg}(i)$  in vacuum, which allows us to determine $\kappa$ since $g$ = 0 in vacuum.  By setting $g$ = 0 in Eq.~(\ref{eq_avrgT}), we get
\begin{equation}\label{eq_avrgT2}
\Delta T_{\rm{avg}}(i)
= \frac{1}{\alpha} \left (\frac{\tan{\beta}}{\beta} - 1 \right ) + \Delta T_{\rm{end}}(i) \, \frac{\tan{\beta}}{\beta} ,
\end{equation}
where
\begin{equation}\label{eq_beta}
\beta \equiv \sqrt{\frac{i^{2} \rho(T_{\rm{amb}}) \alpha L^2}{4 \kappa}}.
\end{equation}
Neglecting the very small $\Delta T_{\rm{end}}(i)$ (less than 0.2 ~K), we can extract $\kappa$, which is listed in Table~\ref{table_conductance}.
We compare the extracted $\kappa$ values with those obtained previously for similar carbon nanotube fibers ($d$ $\sim$ 10~$\mu$m).\cite{BehabtuetAl13Science}   For an acid-doped fiber, $\kappa$ was $\sim$340~W/(m$\cdot$K) at 333~K,\cite{BehabtuetAl13Science} which is consistent with our result, i.e., $\sim$330~W/(m$\cdot$K) at 363~K (average temperature 333~K) for the same diameter fiber.

After determining the value of $\kappa$ from the result obtained in vacuum, the only unknown parameter in $L_{\rm c}$ is $g$; see Eq.~(\ref{eq_critical_length}).  Therefore, we can fit the $\Delta T_{\rm avg}(i)$ curves obtained in three different gases by Eq.~(\ref{eq_avrgT}) with $g$ as the only fitting parameter.  Since $\Delta T_{\rm{end}}$ shows a fairly linear dependence on $I^2$, we use $ \Delta T_{\rm{end}} = \chi I^2$ to determine $ \Delta T_{\rm{end}} (I)$, which we then plug into Eq.~(\ref{eq_avrgT}).  Here, $I = i \pi (d/2)^2$ is the current.  The fitting curves for the fiber in helium is shown in Fig.~\ref{figS1}; see also Fig.~\ref{fig1}(b) for three other gases.

\section{Calculating $g(T)$ based on natural convection theory}

The heat exchange between the fiber and the surrounding gas is governed by the dynamics of gas molecules.  Considering our experimental conditions, we use natural convection theory to analyze this process.  For a horizontal cylinder with $L/d\gg1$, this process can be characterized by the following numbers \cite{Morgan75AHT,GebhartetAl88Book}:
\begin{equation}
\label{eq_convection}
 {\rm Nu}_{d} = \frac{g_{\rm{gas}} d}{\kappa}=0.675 {\rm Ra}_{d}^{0.058},\         10^{-10}\ll {\rm Ra}_{d}\ll 10^{-2}
\end{equation}
\begin{equation}
\label{eq_Ra}
 {\rm Ra}_{d} = \frac{g' \beta (\Delta \overline{T}) d^3}{\nu \alpha'},
\end{equation}
where Nu$_d$ and Ra$_d$ are the Nusselt number and Rayleigh number, respectively, with $d$ as the characteristic length scale; $\kappa$, $\nu$, and $\alpha'$ are the thermal conductivity, kinematic viscosity, and thermal diffusivity of the gas under $1$ atm pressure and at the ``film temperature'' \cite{CengeletAl98Book}:
\begin{equation}
\label{eq_filmT}
T_{\rm f} = {T+T_{\rm{amb}} \over 2}.
\end{equation}
See Table A-10 of Ref.~\onlinecite{CengeletAl98Book} for the values of these parameters.  Here, $\beta$ is the thermal expansion coefficient, equal to $1/T_{\rm f}$ for an ideal gas.

The last two columns of Table~\ref{table_conductance} list the calculated thermal conductance between an ideal cylindrical cable and each individual gas at $\Delta T_{\rm avg}$ = 60~K.  The calculated values are slightly larger than the experimental values.  Equation (\ref{eq_convection}) indicates that $g$ should vary with $d$ as $g \sim d^{-0.826}$, which is close to the experimental results for helium, nitrogen, and argon.

We found no reported $g$ values to compare in this diameter range, while reducing the diameter of CNT bundles to the nanometer range was found to lead to an increase of $g_{\rm air}$ (from 1.5 to 7.9 $\times$ 10$^4$ W/m$\cdot$K).\cite{HsuetAl11JAP}  However, the relative magnitudes of the $g$ values for different gases are quite different from our observation: we find that $g_{\rm{helium}}>g_{\rm{nitrogen}}>g_{\rm{argon}}$ while previous studies found that $g_{\rm{helium}}<g_{\rm{nitrogen}}$.\cite{HuetAl07APL,MannetAl06JPCB} This discrepancy implies the failure of traditional continuum theory of natural convection beyond its scale Ra$_{d} < 10^{-10}$, corresponding to about 1~$\mu$m diameter fiber in air.  More investigations are required to understand the nature of the convection process on sub-micron CNT bundles and how it evolves to larger scale.

\begin{figure*}
\begin{center}
\includegraphics[scale=0.65]{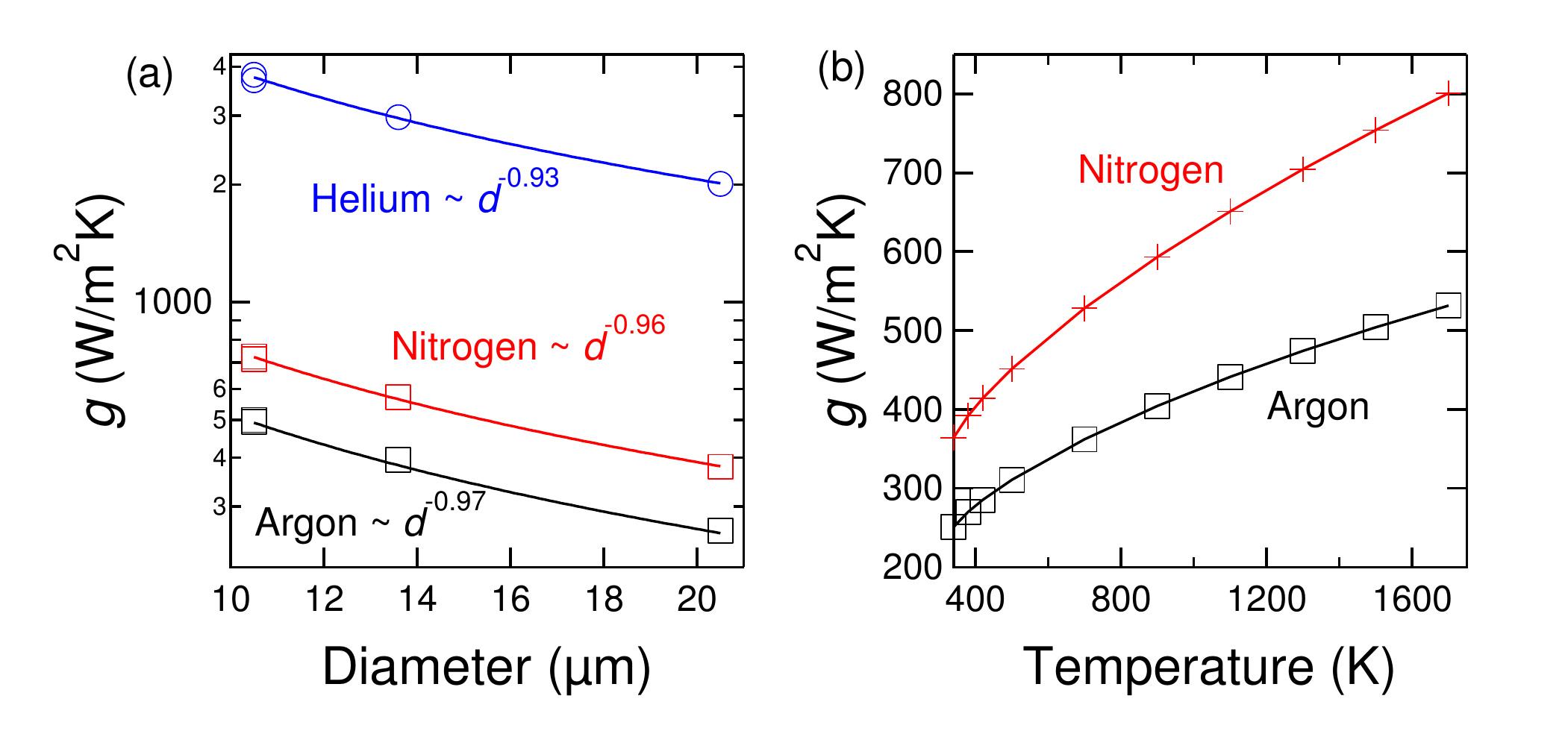}
\caption{(a)~Diameter dependence of $g$ with linear fitting in three different gases.
(b)~Temperature dependence of $g$ for nitrogen and argon based on our measurements and natural convection theory.}
\label{figS2}
\end{center}
\end{figure*}

Finally, we can calculate the temperature dependence of $g$ based on Eqs.~(\ref{eq_convection}) and (\ref{eq_Ra}) using the parameters for nitrogen and argon from a previous study~\cite{HaynesetAl12Book}).  To compensate for the slight difference between the experimental ($g^{\rm exp}$) and theoretical ($g^{\rm theory}$) values at 363~K, we predict the temperature dependence of $g$ to be
\begin{equation}
g^{\rm exp}(T) = \frac{g^{\rm exp}(363 {\rm K})}{g^{\rm theory}(363 {\rm K})} \times g^{\rm theory}(T),
\end{equation}
which is plotted for nitrogen and argon in Fig.~\ref{figS2}.

\section{Estimating the temperatures at the three regime boundaries as well as the FCD of copper}

Irreversible changes start when the maximum temperature of $T(x)$, $T_{\rm max}$, exceeds a certain temperature limit.  If the fiber is infinitely long, then it is homogeneously heated up and any $x$ dependence can be neglected.  Under the stationary condition, $\frac{ \partial T(x,t)}{\partial t}=0$, we get from Eq.~(\ref{eq_heat1})
\begin{equation}
\label{eq_infT_1}
T_{\rm max}=T_{\infty} =T_{\rm amb}+\frac{i^2 \rho(T_{\infty}) d}{4 g (T_{\infty})}.
\end{equation}
We calculated $T_{\infty}$ for the three boundaries by using the corresponding current density values: 1.819 $\times$ 10$^8$, 2.728 $\times$ 10$^8$, and 3.299 $\times$ 10$^8$~A/m$^2$, respectively, from data in nitrogen (1.455 $\times$ 10$^8$, 2.278 $\times$ 10$^8$, and 2.728 $\times$ 10$^8$~A/m$^2$, from the data in argon), which yield 476, 1302, and 1744~K for the three boundaries (463, 1307, and 1736~K from the data in argon).

When $L$ is finite, $T_{\rm max}$ is not equal to $T_{\infty}$ because of the inhomogeneity in $T$, i.e., $T(x)$, imposed by the boundary conditions at the two terminals.  $T(x)$ only varies significantly near the terminals with a length scale given by $L_{\rm c}$ [see Eq.~(\ref{eq_critical_length})].  Although Eq.~(\ref{eq_critical_length}) does not hold for arbitrary temperature ranges, it can provide insight into how $L_{\rm c}$ depends on other temperature-dependent parameters.  First, $L_{\rm c}$ decreases when $g/\kappa$ increases because the heat dissipation into the terminals (characterized by $\kappa$) relative to the gas (characterized by $g$) decreases.  The factor $i^2 \rho(T_{\rm{amb}})\alpha$ in Eq.~(\ref{eq_critical_length}) can be rewritten as $i^2 \Delta \rho_{\rm avg} /\Delta T_{\rm avg}$, which reveals a small correction due to the temperature-dependent resistivity. Hence, in general, $L_{\rm c}$ can be written as $L_{c}^2 \sim g/\kappa+\Delta (g/\kappa)+i^2 \Delta  \rho_{\rm avg} /\Delta T_{\rm avg}$.  Above $T_{\rm amb}$, $\kappa(T)$ decreases with increasing $T$, exhibiting $\kappa \propto 1/T$ at high temperatures due to Umklapp phonon scattering.  $g$ also increases with $T$ (see the previous section). Meanwhile, we can estimate that the correction due to $\Delta \rho_{\rm avg}$ is negligible (at most $10\%$ compared with the first term).  Therefore, $L_{\rm c}$ should decrease with increasing $T$, making the difference between $T_{\rm max}$ and $T_{\infty}$ smaller at high temperatures.  
In Regime 1, $\Delta T(x)$ is still small and $\rho(T)$ does not deviate much from a linear function (we measured $\rho(T)$ up to 423~K, where $\rho(T)$ was still quiet linear).  If we regard $\kappa$ and $g$ as constants, Eq.~(\ref{eq_temp}) still applies:
\begin{equation}
\label{eq_temp_max}
  \Delta T_{\rm max} = \left( 1-\frac{1}{\cosh(L/2L_{\rm c})}\right)\Delta T_{\infty}+\Delta \overline{T}_{\rm{BC}} \frac{1}{\cosh(L/2L_{\rm c})}.
\end{equation}
By using the maximum value of $\kappa$, i.e., $\kappa(T_{\rm amb})$, together with $g_{\rm nitrogen} (\frac{T_{amb}+T_{\infty}}{2})$, the lower limit of $L/L_{\rm c}$ is calculated to be about 15 for a 20-$\mu$m fiber, and  $\Delta T_{\rm max} \sim 0.998\ \Delta T_{\infty}$. So we may conclude that through the whole current range of our experiment, $ T_{\rm max} \sim T_{\infty}$. The same argument is also valid for a copper wire with 20~$\mu$m diameter.  To estimate the FCD of copper, $\Delta T_{\rm max} $ is set to be the melting temperature of copper (1357.6~K), and we obtain $\rho^{\rm copper}_{\rm FCD}$ = 10.171 $\times$ 10$^{-8}$~$\Omega$/m.\cite{Matula79JPCRD}

\section{Comprehensive comparison of MCDBB among buckypapers, carbon fibers, CNT fibers, and CNT fibrils}

\begin{table*}
  \caption{MCDBB of buckypapers, carbon fibers, CNT fibers, and CNT fibrils.  The MCDBB for the acid-doped CNT fiber with $d$ = 4.2~$\mu$m was calculated based on the value of the fiber with $d$ = 10.5~$\mu$m. All other listed values are from experiments.}
  \label{table_MCDBB}
  \begin{center}
  \begin{tabular}{lcccc}
    \hline
    \hline
     & $d$ & Test condition & MCDBB & Reference \\
     & ($\mu$m) & & ($10^7$~A/m$^2$) &   \\
     \hline
     Acid-doped CNT fiber & 20.5 & in air & 2.11 & this work\\
                          & 20.5 & in vacuum & 1.36 &  \\
    \hline
     Acid-doped CNT fiber & 10.5 & on substrate & 115 to 144 & this work  \\
                          & 4.2 & on substrate & 162 to 225 &  \\
     \hline
     Bulkypaper & 20 & in air & 0.11 & Ref.~\onlinecite{ParketAl08Nanotech} \\
                & 20 & in vacuum & 0.31 & \\
     \hline
     CNT fiber & 13.3 & on substrate & 42.7 & Ref.~\onlinecite{SongetAl12Nanotechnology} \\
               & 9.1 & on substrate & 66.4 &  \\
               & 5.6 & on substrate & 103 &  \\
     \hline
     Carbon fiber &6.1& on substrate & 82.1 & Ref.~\onlinecite{SongetAl12Nanotechnology} \\
     \hline
     CNT fiber & 4.2 & on substrate & 162 & Ref.~\onlinecite{ZhaoetAl11SciRep} \\
     \hline
     \hline
  \end{tabular}
  \end{center}
\end{table*}

Both the FCD and MCDBB depend on the dimensions of the cable as well as the surrounding thermal media. In Table 1 of Ref.~\onlinecite{ParketAl08Nanotech}, the values of the breakdown current density (the same as the MCDBB in this paper) for several suspended buckypapers in air and vacuum are listed.  The largest value corresponds to the ``A-BP'' sample, which is 20~$\mu$m thick and 20~cm long; these dimensions are comparable to those of our fibers here (20~$\mu$m in diameter and 30~cm in length).  The MCDBB for this sample is 1.1~$\times$~10$^6$~A/m$^2$ in air and 3.1~$\times$~10$^6$~A/m$^2$ in vacuum, both of which are much smaller than the values for our acid-doped fiber (2.11~$\times$~10$^7$~A/m$^2$ in air and 1.36~$\times$~10$^7$~A/m$^2$ in vacuum).

In Table 1 of Ref.~\onlinecite{SongetAl12Nanotechnology}, the maximum current values of several CNT fibers as well as a carbon fiber, all laid on substrates, are listed. The lengths of those samples were only about 500~$\mu$m, and therefore, heat dissipation is not only through the substrate but also through the electrodes (heat dissipation through the air can be ignored). Since it is very hard to make identical solid-to-solid interfaces between a fiber and those two media in each test, the measured MCDBB values would contain large test-to-test fluctuations. Nevertheless, we tested several acid-doped fibers with 10~$\mu$m diameter and $\sim$1~mm length on substrates; the maximum current varied from 100 to 125~mA, corresponding to 1.15~$\times$~10$^9$~A/m$^2$ to 1.44~$\times$~10$^9$~A/m$^2$ in MCDBB, more than 2 times the value of the fiber with either 9.1 or 13.3~$\mu$m diameter in Ref.~\onlinecite{SongetAl12Nanotechnology}. It is also larger than the MCDBB of a PAN carbon fiber with 6.1~$\mu$m diameter (0.82~$\times$~10$^9$~A/m$^2$) in Ref.~\onlinecite{SongetAl12Nanotechnology}. It even exceeds the largest MCDBB (1.03~$\times$~10$^9$~A/m$^2$) in Ref.~\onlinecite{SongetAl12Nanotechnology}, which was for the smallest fibril with 5.6~$\mu$m diameter.

Reducing the diameter of the fiber is expected to improve the FCD and MCDBB values because of the enlarged surface-to-volume ratio.  Namely, the amount of Joule heating is $\propto i^2 d^2 L$, which is proportional to the volume, while heat dissipation is $\propto d L$, which is proportional to the surface area.  Therefore, the maximum current density, which is determined by the balance between the two, should be proportional to $d^{-0.5}$ (assuming that the other parameters do not vary with the diameter).  Based on this expectation, we can project a factor of 1.35 improvement in MCDBB when the diameter of the acid-doped fiber decreases from 10.5 to 5.6~$\mu$m.  Similarly, we can estimate the MCDBB of acid-doped fibers with diameters ranging from 10.5 to 4.2~$\mu$m to compare with the fiber reported in the Supplementary Information of Ref.~\onlinecite{ZhaoetAl11SciRep} (although the length of the fiber was not specified).  The projected MCDBB varies from 1.62~$\times$~10$^9$ to 2.25~$\times$~10$^9$~A/m$^2$, which are better than the value of 1.62~$\times$~10$^9$~A/m$^2$ reported in Ref.~\onlinecite{ZhaoetAl11SciRep}.

\section{Effect of annealing on the Raman G/D ratio}

\begin{figure}
\begin{center}
\includegraphics[scale=0.475]{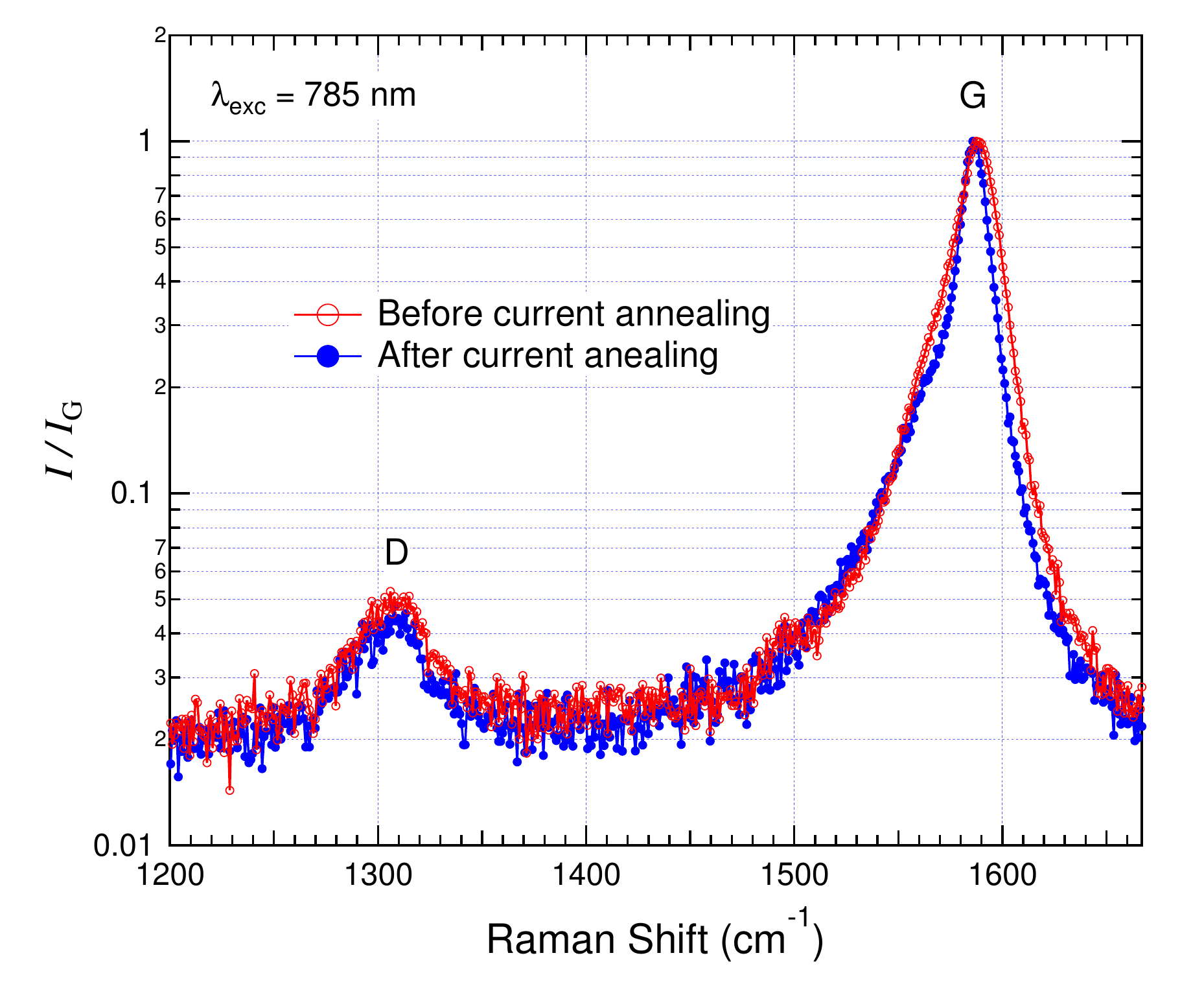}
\caption{Raman spectra for a carbon nanotube fiber taken before and after heating with high current, corresponding to Regime 2 of Figure 1c of the main text.  The very high G/D peak intensity ratio is well maintained between the two spectra, indicating that the heating process did not damage the fiber.}
\label{figS3}
\end{center}
\end{figure}

To demonstrate that the quality of the fiber is not degraded during the irreversible current-heating process in Regime 2, we performed Raman spectroscopy before and after going through the heating process.  Figure~\ref{figS3} shows that the G/D ratio is very well maintained at a small value (4-5 $\times$ 10$^{-2}$).

\end{document}